\documentstyle[12pt]{article}

\begin{document}

\title{Superkamiokande and solar antineutrinos}
\author{G.Fiorentini$^{1,2}$, M.Moretti$^{1,2}$ and F.L.Villante$^{1,2}$}

\date{July 8, 1997}

\maketitle

\small{$^{1}$Dipartimento di Fisica, Universit\`a di Ferrara, I-44100,
Ferrara (Italy)\\
$^{2}$Istituto Nazionale di Fisica Nucleare, Sez. di Ferrara,
I-44100, Ferrara (Italy)}
\bigskip
\bigskip

\begin {abstract}

We propose to exploit the angular distribution of 
 the positrons emitted in the reaction
$\overline{\nu}_{e}+p\rightarrow n+e^{+}$ 
to extract a possible antineutrino signal from
the Superkamiokande background. From the statistics collected in 
just 101.9 days one obtains a model independent upper bound 
on the antineutrino flux
$\Phi_{\overline{\nu}}(E_{\overline{\nu}}>8.3MeV)<9\cdot 10^{4}
 cm^{-2}s^{-1}$
at the 95\% C.L. By assuming the same energy spectrum as for the $^{8}$B
neutrinos, the 95\% C.L. bound is 
$\Phi_{\overline{\nu}}(E_{\overline{\nu}}>8.3MeV)<6\cdot 10^{4}
 cm^{-2}s^{-1}$.
Within three years of data
taking, the sensitivity to $\nu_{e}\rightarrow\overline{\nu}_{e}$
 transition probability will reach the 1\%
level, thus providing a 
 stringent 
test of hybrid oscillation models.
\end{abstract}

\pagebreak

We all believe that sun light is accompanied by an intense
neutrino radiation ($\Phi_{\nu}\sim10^{11}cm^{-2}s^{-1}$,
i.e. about one neutrino per $10^{7}$ photons). Since energy is
produced by transforming hydrogen into helium, the conservation
of electric charge and
lepton number requires that this radiation consists of
neutrinos and not of antineutrinos, see e.g.~\cite{ba1,ber,cas} for
reviews on solar neutrinos. 
However, a fraction of the $\nu_{e}$
formed in the core of the Sun could transform into $\overline{\nu}_{e}$
during their trip from Sun to Earth; this transformation
is predicted e.g.
in the so called hybrid models \cite{ak1,ak2,lim} 
 where a spin-flavour
magnetic moment transition gives $\nu^{L}_{e}\rightarrow
\overline{\nu}^{R}_{\mu}$ and a mass oscillation
 yields $\overline\nu^{R}_{\mu}\rightarrow\overline{\nu}^{R}_{e}$.
 Solar antineutrinos could also originate from neutrino
decay \cite {cab} both in vacuum \cite{dec} and in matter \cite{zur}.
 All this shows that an experimental study of solar antineutrinos
is important.

As well known, the specific signature of antineutrinos in hydrogen
containing materials is through the inverse beta decay (I$\beta$D),
$\overline{\nu}_{e}+p\rightarrow n+e^{+}$,
which produces {\it almost isotropically} distributed monoenergetic 
positrons ($E_{e^{+}}=E_{\overline{\nu}}-\Delta m; \Delta m=m_{n}-m_{p}$);
for energy above a few MeV, the cross section
is:

\begin{equation}
\sigma_{0}(E_{\overline{\nu}})=9.2\times10^{-42} cm^{2}
[(E_{\overline{\nu}}-\Delta m)/10 MeV]^{2}
\end {equation}

\par
Water \u{C}erenkov detectors as Kamiokande \cite{kam,kam2,kam3,kam4}
 and the recently
operational Superkamiokande (SK)
 \cite{skam}, which are sensible to the
 $\nu_{e}-e$
interaction with a much smaller cross section, are clearly also capable
of detecting I$\beta$D, which produces a number $N_+$ of events given by:

\begin{equation}
N_{+}=N_{p}T\epsilon\Phi_{\overline{\nu}}(E_{\overline{\nu}}>E_{o})
\overline{\sigma}_{0}
\end{equation}

\noindent
where $N_{p}$ is the number of free protons, T is the exposure time,
 $\epsilon$
is the (assumed constant) detection efficiency, $\Phi_{\overline{\nu}}$ is
 the 
antineutrino flux, $E_{0}$ the minimal detectable antineutrino energy
and $\overline{\sigma}_{0}$ is the
cross section averaged over the antineutrino spectrum for
$E_{\overline{\nu}}>E_{0}$:

\begin{equation}
\overline{\sigma}_{0}=\frac{\int_{E_{0}}^{\infty}
dE_{\overline{\nu}}\,\sigma_{0}(E_{\overline{\nu}})
w(E_{\overline{\nu}})}{\int_{E_{0}}^{\infty}
dE_{\overline{\nu}}\,w(E_{\overline{\nu}})}
\end{equation}

\noindent
where $w$ is the antineutrino probability distribution. 
\par
 Antineutrino events contribute to the isotropic background $B$. By requiring
 $N_{+}<B$, upper bounds on $\Phi_{\overline{\nu}}$ have been derived from
 Kamiokande data
 \cite{bar}: for $E_{\overline{\nu}} \geq 9.3 MeV$ the antineutrino
 flux does not exceed $6-10\%$ of Standard
 Solar Model (SSM) predictions for the $\nu_{e}$-neutrino flux in the same
energy range. 
A similar result
 can be obtained by SK, as the ratio $B/(N_{p}T)$ is similar for
 the two detectors.\footnote{One should notice that the rough
 equality of $B/(N_{p}T)$ in the two detectors occurs although the
SK energy threshold is significantly lower
than that of Kamiokande.
This is to say that the quality of SK data has neatly improved
so that lower energies can now be explored 
 \cite{bar}.}

\par
The aim of this letter is to show that a much better sensitivity can
be achieved by exploiting the huge statistics of SK in conjunction
with the (although weak) {\it directionality} of positrons from I$\beta$D. In
fact, a signature for the presence of positrons from antineutrinos is
provided by the angular dependence of the cross section:

\begin{equation}
\frac{d\sigma}{d\cos\theta}=\frac{\sigma_0}{2}(1-a\cos\theta)
\label {angdep}
\end{equation}

\noindent
where

\begin{equation}
 a=\frac{(g_{A}/g_{V})^{2}-1}{3(g_{A}/g_{V})^{2}+1}\simeq0.1
\end{equation}

\noindent
and  $g_{V,A}$ are respectively the vector and axial
 couplings of the neutron.

In the angular region where events from the $\nu-e$ interactions
can be neglected
(see fig.1),
 a linear fit to the counting yield, $C=C_{0}-C_{1}\cos\theta$, gives the
 slope $C_{1}$ and thus the antineutrino flux through the relation:

\begin{equation}
\Phi_{\overline{\nu}}(E_{\overline{\nu}}>E_{0})=
\frac{C_{1}}{T}\frac{2}{N_{p}\epsilon\overline{\sigma}_{0}a}
\end{equation}

\par
We remark the advantages of this method with respect to the previous
one:
\begin{itemize}
\item
The isotropic background 
can be subtracted.
\item
The determination of $C_{1}$ provides a mean for {\it detecting}
antineutrinos from the Sun (and not only for deriving upper bounds).
\item
The sensitivity to antineutrinos increases as statistics accumulates.
In fact the accuracy on the slope $C_{1}$ is limited by statistical
fluctuations, $\Delta C_{1}\sim\sqrt{B}\propto\sqrt{N_{P}T}$.
and consequentely $\Delta\Phi_{\overline{\nu}}\propto 1/\sqrt{N_{p}T}$.
 On the other hand, in the previous method there is no gain in increasing
statistics as the ratio $B/N_{P}T$ stays constant.
\end{itemize}

\par
\bigskip
In order to provide a quantitative illustration
of the previous points
we used data from the first 101.9 operational days of SK, as reported in 
fig.3 of \cite{tot}. 
By considering the region $-1<cos\theta<0.5$ and taking into account
the finite angular resolution, after subtracting the tail due the
$\nu-e$ scattering, as a result of this exercise we obtain:
\begin{equation}
\frac{C_{1}}{T}=(-0.7\pm1.5) day^{-1}
\end{equation}
\par
\bigskip
Let us remind that an antineutrino signal requires $C_{1}$ to
 be positive;
 with
this constraint, following the prescription of \cite{par} one has $C_{1}/T
<2.5 \, day^{-1}$ to the 95\% C.L.

\par
We now proceed to extract a bound on the antineutrino flux.\footnote
{For semplicity we neglect the finite energy resolution of the detector,
which should be considered in a proper analysis.} We assume
$\epsilon=0.95$ \cite{tot} for visible energy $E_{vis}\geq7MeV$;
 since the visible energy can be identified with the total
electron/positron energy, the minimal antineutrino energy is 
$E_{0}=8.3MeV$.
 In order to determine
 $\overline{\sigma}_{0}$ we consider two approches:

\smallskip
\par\noindent
$a)$ If {\it one assumes} that the antineutrino spectrum has
 the same
 shape as that of $^{8}$B solar neutrinos, one has
 $\overline{\sigma}_{0}=7.06\cdot10^{-42}cm^{2}$. 
This gives $\Phi_{\overline{\nu}}(E_{\overline{\nu}}>8.3MeV)<6
 \cdot 10^{4} cm^{-2}s^{-1}$,
 to the 95\% C.L.
 This bound corresponds to a fraction x=3.5\% of the solar neutrino flux
 (in the energy range $E_{\nu}>8.3MeV$) predicted by the SSM \cite{ba2}.
\smallskip
\par\noindent
$b)$ Alternatively one get a model independent bound by releasing
the assumption on the antineutrino energy spectrum. As $\sigma_{0}$
increases with $E_{\overline{\nu}}$, clearly
$\overline{\sigma}_{0}\geq\sigma_{0}(E_{0})=4.5\cdot10^{-42}cm^{2}$.
This gives a {\it model independent} bound $\Phi_{\overline{\nu}}
(E_{\overline{\nu}}>8.3MeV)<9\cdot 10^{4} cm^{-2}s^{-1}$ to the 95\% C.L.
\bigskip

 An additional way to  discriminate a solar antineutrino signal
from "true background" (which should be time independent)
is provided by  the study  of seasonal
effects, due to the variations of the Sun-Earth distance. The time
distribution of positrons from  $I\beta D$, to the first order
in the eccentricity $e$ of the Earth orbit ($e=0.0168$), is given by:
\begin{equation}
\frac {d N_+ } {dt} = N_{P}\epsilon
\overline{\sigma}_{0}\Phi_{\overline{\nu}}[ 1 + 2e\cos(\omega t)], 
\label{season}
\end{equation}
where
$\Phi_{\overline{\nu}}$ is the yearly averaged antineutrino flux,
 $t$ is the time ($t=0$ at the perihelion )
 and obviously $\omega= 2\pi \ {\rm years}^{-1}$. 
As the
amplitudes of the oscillating component in (\ref{season})
 and in (\ref{angdep}) are comparable,
also this method will allow, in due time, an accurate measurement
of the solar antineutrino flux: quantitatively, the minimum detectable
flux  will be a 35\% higher than that of the previous method. Both approaches
-- angular correlation and seasonal effects -- clearly should be studied,
as being independent from and complementary to each other.

\bigskip
\par
In summary:

\smallskip
\begin{itemize}
\item
The (although weak) directionality of inverse beta decay, combined with
the huge statistics of SK allows for the search of the solar antineutrinos
(and not only the derivation of upper bounds).

\item
By using this method and the first available data on SK one already
improves on the information provided by Kamiokande \cite{bar} (where the 
statistics was too low for the new method to be useful).

\item 
We would like to encourage our experimental colleagues to analyse 
in this spirit the available data and those which will be collected
 in the future. Within three years
of data taking, the sensitivity
 to $\nu_{e}\rightarrow\overline{\nu}_{e}$
transition probability will reach the 1\% level, thus allowing for
a definite test of hybrid oscillation models. One could also detect
galactic antineutrino sources with luminosities
$L_{\overline{\nu}}\geq 3\cdot10^{45}erg/s$ (i.e $L_{\overline{\nu}}
\sim10^{12}L_{\odot}$), should they exist. 
\end{itemize}
%%%%%%%%%%%%%%%%%%%%%%%%%%%%%%%%%%%%%%%%%%%%%%%%%%%%%

%%%%%%%%%%%%%%%%%%%%%%%%%%%%%%%%%%%%%%%%%%%%%%%%%%%%%%

\pagebreak

\begin{center}
{\bf Figure caption}
\end{center}
\bigskip
\par\noindent
Sketch of the expected angular distribution of events in the presence
of a solar $\overline{\nu}_{e}$ flux
   
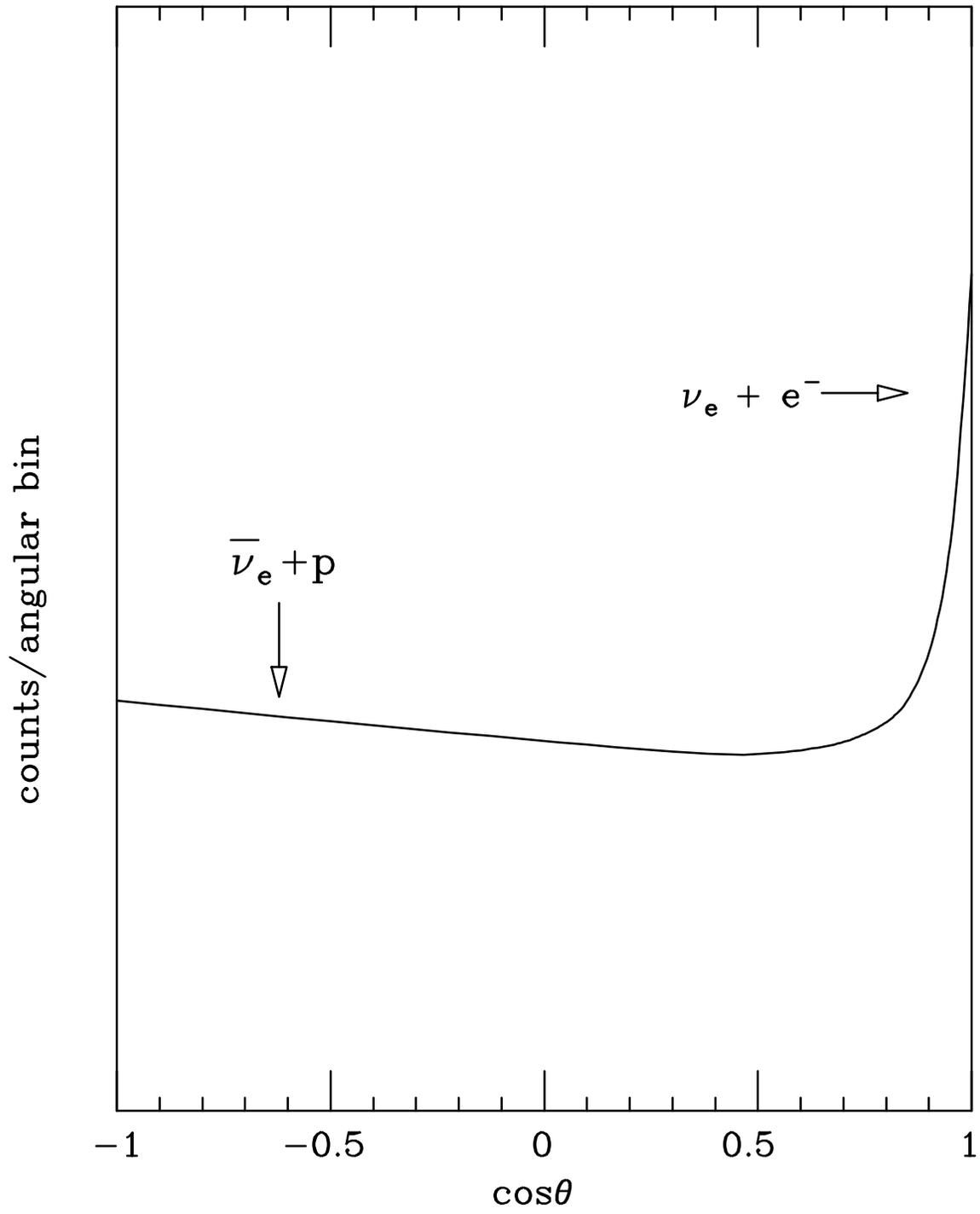
\end{document}